\documentclass{article}
\usepackage{amsmath,graphicx,subfig}
\usepackage{amssymb,amsbsy}
\usepackage{tikz}
\author{Adel Khoudeir and J. Stephany}
\title{A non local unitary vector model  in 3-D}

\begin{document}
\tikzstyle{bag} = [text width=2em, text centered]
\tikzstyle{end} = []

\begin{titlepage}

\hfill{ \parbox{3cm}{{AEI-2011-061}\newline {SB/F/388-11}}} 
\hrule \vskip 2cm

\begin{center}{\large{\bf A non local unitary vector model  in 3-D }}
\vskip1cm
Adel Khoudeir \\ 
\vskip 0.2cm
{\it Centro de F\'{\i}sica Fundamental, Departamento de F\'{\i}sica, \\
Facultad de Ciencias, Universidad de Los Andes, M\'erida 5101, Venezuela\\
adel@ula.ve}
\\ 
\vskip 0.3cm
J. Stephany\\
\vskip 0.2cm
{\it Max-Planck-Institut f\"ur Gravitationsphysik,Albert-Einstein-Institut\\
Am M\"uhlenberg 1, 14476 Golm, Germany}\\
{\it and}\\
{\it Departamento de F\'{\i}sica, Universidad Sim\'{o}n Bol\'{\i}var, \\
Apartado Postal 89000, Caracas 1080-A, Venezuela\\
stephany@usb.ve}\\

\end{center}
\vskip 1cm
\begin{abstract}
We present a unified analysis of single excitation vector models in $3D$. We show that there is a
family of first order master actions related by duality transformations which interpolate between the different models. We use a Hamiltonian (2+1) analysis to show the equivalence of the self-dual and topologically massive models with a covariant non local  model which  propagates also a single massive excitation. It is  shown how the non local terms appears naturally  in the path integral framework.
\end{abstract}
\vskip 2cm
\hrule
\bigskip
\vfill
\end{titlepage}

\section{Introduction} 

Three dimensional vector and gravity models present the most simple set up where
duality transformations allow an explicit equivalence of
different systems. They also have intrinsic interest for the particular
mechanism which generates the mass of the excitations by the incorporation of topological
terms \cite{DesSJT1982,SchJ1981}. Both these facts have been present in the
recent discussions of alternative higher order actions for three dimensional
massive gravity \cite{BerEHT2009,DalDM2009,DalDM2010,AriPS2011} which generalize the
Fierz-Pauli theory and the parity sensible models.  These models provide
exceptions \cite{DesS2010} to the standard  association of higher order actions
with ghost propagation generally expected inboth  the bosonic \cite{SteK1977}
and the fermionic \cite{AraCS1986a} cases. This should prompt  interest to
investigate the limitations of that otherwise very useful guiding principle  as
already proposed in Ref.\cite{BenCM2008} in a different context.  The form in
which the intermediate master actions appear in the analysis of vector and gravity 
models in $3D$ suggests a mechanism for generating  models with higher
derivatives which may be unitary.  In this paper we consider  this
possibility for the case of vector fields and construct with this goal, a
hierarchy of master actions. We show that the mechanism which generates the
higher order equations  saturates in this case at the second order. Nevertheless  instead of a
third order unitary model, we find that a non local unitary model which describes a
single massive excitation appears. This is interesting since non-local models, which were introduced early
in quantum field theory \cite{PauW1953,ChrMP1954} and play an important role through the effective action in the functional approach \cite{JacR1974,JacRP1997}, have been found in recent years to be of importance in defining the dynamics of higher spin fields \cite{FraDS2002}.

\section{Vector models in 3-D}\label{Vector}

Local, covariant unitary massive vector models in $3D$ may describe either one  or two
excitations with definite parity. The usual Proca model (PM) with a  Fierz-Pauli mass term whose action is
\begin{equation}\label{FP}
  I_{FP}[A]=<-\frac{1}{4}F^{\mu\nu}(A)F_{\mu\nu}(A)-\frac{m}{2}A_\mu A^\mu> \ .
\end{equation}
describes two parity sensible excitations. The self dual
model (SDM)\cite{TowPPV1984} 
\begin{equation}\label{Selfdual}
   I_0[a]=<-\frac{m}{2}a_\mu a^\mu+\frac{1}{2}a_\mu
\epsilon^{\mu\nu\rho}\partial_\nu a_\rho> \ .
\end{equation}
and the topologically massive model (TMM)\cite{DesSJT1982,SchJ1981}  
\begin{equation}\label{TMM}
   I^{TM} [A]=<\frac{1}{2m}\epsilon^{\mu\nu\rho}\partial_\nu
A_\rho\epsilon_{\mu\sigma\tau}\partial^\sigma A^\tau-\frac{1}{2} A_\mu
\epsilon^{\mu\nu\rho}\partial_\nu A_\rho> .
\end{equation}
describe each a single excitation of definite parity.

Locally, these last two models are canonically equivalent \cite{DesSJ1984}, the SDM
being a gauge fixed representation of the gauge invariant  TMM
\cite{GiaRRS1991,AriPS1995,RestAS1993,BanRRR1997}. The TMM and the SDM are also related by a duality transformation. This equivalence is most
compactly encoded in the master action \cite{DesSJ1984} which allows to connect
them in a covariant way. It is given by 
\begin{equation}\label{Master}
   I_1 [a, A]=<-\frac{m}{2}a_\mu a^\mu+A_\mu
\epsilon^{\mu\nu\rho}\partial_\nu a_\rho -\frac{1}{2} A_\mu
\epsilon^{\mu\nu\rho}\partial_\nu A_\rho> .
\end{equation}
Taking variations with respect to $A_\mu$ one obtains the identity 
\begin{equation}\label{Rotor}
\epsilon^{\mu\nu\rho}\partial_\nu a_\rho=\epsilon^{\mu\nu\rho}\partial_\nu
A_\rho \ \ ,
\end{equation}
which assures  that the transverse parts of $A_\mu$ and $a_\mu$ are equal.
Substitution in the master action leads to the self dual action $
I_1[a,A(a)]=I_0[a_\mu]$. 
This procedure may be justified using the gauge invariance of the action.
On the other hand, the equation which results from taking variations with respect to
$a_\mu$ establishes that $a_\mu$ is transverse. We should write it in the form
\begin{equation}\label{amu}
  a_\mu(A)=\frac{1}{m}\epsilon^{\mu\nu\rho}\partial_\nu A_\rho \ \ ,
\end{equation}
stressing its structure as a kind of  change of variables. Note that for $a=A$ this is the equation of motion of the SDM. Upon substitution in (\ref{Master})  the second order gauge
invariant TMM action (\ref{TMM}) is generated.
Since both (\ref{Rotor}) and (\ref{amu}) involve time derivatives one may in
principle wonder if this procedure guarantees canonical equivalence.  In this
case  it is indeed true that the two models  (\ref{Selfdual}) and (\ref{TMM})
are canonically equivalent  but, as we discuss below, a similar strategy may in
other cases  connect non-equivalent models.  

In order to search for higher order models or alternative formulations of the TMM 
and the SDM, one may use again the trick relating $I_1[a,A]$ with $I_0[a,]$ and  introduce a
second intermediate equivalent action given by,
\begin{equation}\label{I2}
   I_2 [a, A, B]=<-\frac{m}{2}a_\mu a^\mu+a_\mu
\epsilon^{\mu\nu\rho}\partial_\nu A_\rho -A_\mu
\epsilon^{\mu\nu\rho}\partial_\nu B_\rho+\frac{1}{2} B_\mu
\epsilon^{\mu\nu\rho}\partial_\nu B_\rho> \ ,
\end{equation}
The  equation of motion obtained by taking variations with respect to $B$ relates $A$
and $B$ in the same form as (\ref{Rotor}) relates $a$ and $A$ in $I_1$. Thus taking into account the gauge invariance, $I_2[a,A,B]$ is also equivalent to $I_1[a,A]$. 
If instead we use (\ref{amu}) which is again the equation obtained by taking
variations with respect to $a_\mu$, we end up with 
\begin{equation}\label{TMM2}
   I^{TM}_2[A,B ]=<\frac{1}{2m}\epsilon^{\mu\nu\rho}\partial_\nu
A_\rho\epsilon_{\mu\sigma\tau}\partial^\sigma A^\tau-A_\mu
\epsilon^{\mu\nu\rho}\partial_\nu B_\rho+\frac{1}{2} B_\mu
\epsilon^{\mu\nu\rho}\partial_\nu B_\rho>  .
\end{equation}
By the discussion just presented, which will be complemented by the Hamiltonian analysis of section (\ref{Hamiltonian})
this action is also equivalent to $I_{TM}$. As
discussed in Refs. \cite{SteJ1997,LeGJMN1997} the introduction of these new  fields
corresponds to a duality transformation done in the quantum mechanical
generating functional. Although the different models  are seen to be locally
equivalent they are not equivalent on topologically non trivial manifolds
\cite{AriPR1996}. In the path integral formulation this is reflected in that the
generating functionals of the models differ by a factor which is a power of the
pure Chern Simons generating functional. We  also note that an explicit Chern
Simons term $<-\frac{\beta}{2}A_\mu \epsilon^{\mu \nu \rho}\partial_\nu A_\rho>$ for the
field $A$ may be included in (\ref{I2}) or (\ref{TMM2}) without affecting the
symmetries of the action. In fact it can be seen that this amounts to a shift of
the fields and a redefinition of the mass.

We can generalize this procedure and introduce  a family of equivalent first
order actions 
\begin{eqnarray}\label{IN}
   I_N [A^J]=<-\frac{m}{2}A^{(0)}_\mu A^{(0)\mu}+A^{(0)}_\mu
\epsilon^{\mu\nu\rho}\partial_\nu A^{(1)}_\rho -A^{(1)}_\mu
\epsilon^{\mu\nu\rho}\partial_\nu A^{(2)}_\rho+
\nonumber\\\cdots + (-1)^N\frac{1}{2} A^{(N)}_\mu
\epsilon^{\mu\nu\rho}\partial_\nu A^{(N)}_\rho>\ \ ,
\end{eqnarray}
with alternating signs for the coupling terms and $N\ge1,\ \ J\epsilon [0,N]$. We can also construct
a family of actions directly equivalent to $I_{TM}^2$  through the same
mechanism.
\begin{eqnarray}\label{TMMN}
   I^{TM}_N [A^J]&=&<\frac{1}{2m}\epsilon^{\mu\nu\rho}\partial_\nu
A^{(1)}_\rho\epsilon_{\mu\sigma\tau}\partial^\sigma A^{(1)\tau}-A^{(1)}_\mu
\epsilon^{\mu\nu\rho}\partial_\nu A^{(2)}_\rho  \nonumber\\ 
&+& A^{(2)}_\mu \epsilon^{\mu\nu\rho}\partial_\nu A^{(3)}_\rho+
\cdots + (-1)^N\frac{1}{2} A^{(N)}_\mu \epsilon^{\mu\nu\rho}\partial_\nu
A^{(N)}_\rho> \  .
\end{eqnarray}
with $N\ge 2,\ \ J\epsilon [1,N]$. Each of these families is generated by applying iterate duality
transformations either to the SDM  or the TMM.

\section{Models with higher derivatives}\label{Higher}

\begin{figure}[ht]
\begin{center}
\begin{tikzpicture}[sloped]
  \node (a) at ( 0,0) [bag] {$I_3$};

  \node (b) at ( 2,1) [bag] {$I_2$};
  \node (c) at ( 2,-1) [bag] {$I_3^{TM}$};

  \draw [->] (a) to node [above] {} (b);
  \draw [->] (a) to node [below] {} (c);

  \node (d) at ( 4,2) [bag] {$I_1$};
  \node (e) at ( 4,0) [bag] {$I_2^{TM}$};
  \node (f) at ( 4,-2) [bag] {$I^{3th}_2$};

  \draw [dotted,->] (c) to node [above] {} (f);
  \draw [->] (c) to node [below] {} (e);
  \draw [->] (b) to node [above] {} (e);
  \draw [->] (b) to node [below] {} (d);

  \node (g) at ( 6,3) [bag] {$I_0$};
  \node (h) at ( 6,1)   [bag] {$I^{TM}$};
  \node (i) at ( 6,-1) [bag] {$I^{3th}$};
  \node (j) at ( 6,-3) [bag] {$I^{4th}$};
  \node (k) at ( 8,-2) [bag] {$I^{4th}_{FP}$};
\node (l) at ( 8,0) [bag] {$I^{3th}_{FP}$};

  \draw [->] (d) to node [above] {} (g);
  \draw [->] (d) to node [below] {} (h);
  \draw [->] (e) to node [above] {} (h);
  \draw [dotted,->] (e) to node [below] {} (i);
  \draw [dotted,->] (f) to node [above] {} (i);
  \draw [dotted,->] (k) to node [below] {} (i);
\draw [dotted,->] (l) to node [below] {} (i);
\draw [dotted,->] (k) to node [below] {} (j);
 \draw[-latex,color=red]
        (g) .. controls +(right:.2cm) and
                                +(right:3cm) ..
            node[near end,above right,color=black] {}
        (i);
\draw[-latex,color=blue]
        (h) .. controls +(left:.2cm) and
                                +(left:3cm) ..
            node[near end,above right,color=black] {}
        (j);
\draw[-latex,color=green]
        (d) .. controls +(left:.2cm) and
                                +(left:3cm) ..
            node[near end,above right,color=black] {}
        (f);
\draw[-latex,color=magenta]
        (d) .. controls +(right:.2cm) and
                                +(right:3cm) ..
            node[near end,above right,color=black] {}
        (k);
\end{tikzpicture}
\end{center}
\caption{Vector models with a single excitation}
\label{tree}
\end{figure}
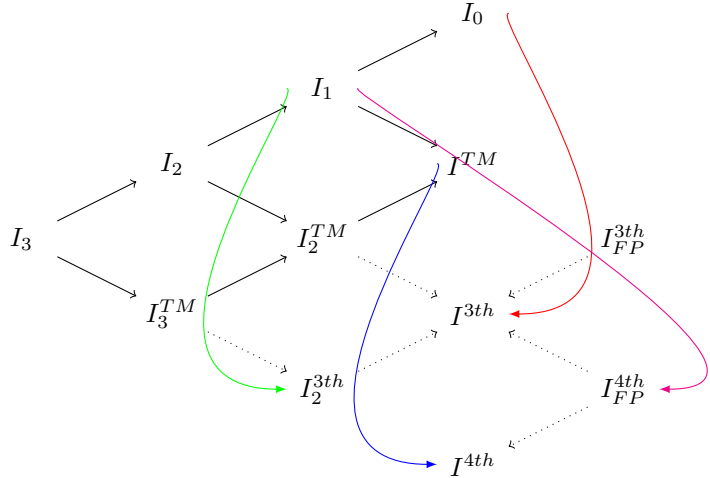

The actions considered till now and others which appear in the discussion below are arranged in Fig.\ref{tree}. Straight arrows connect physically equivalent actions \noindent and  dashed arrows denote  relations between actions when at least one of them is not unitary. To continue, we observe that in the same way that $I_1$ leads to the second order
topologically massive model  $I^{TM}$, the action  $I_2$ should generate a third
order model  $I^{3th}$. 
Substituting the equation of motion 
\begin{equation}\label{EMTMM2}
 \epsilon^{\mu\nu\rho}\partial_\nu B_\rho=
\frac{1}{m}\epsilon^{\mu\nu\rho}\partial_\nu
\epsilon_{\rho\sigma\tau}\partial^\sigma A_\tau 
\end{equation}
in $I^{TM}_2$ one obtains
\begin{equation}\label{3th}
 I^{3th}[A]=<\frac{1}{2m^2}\epsilon^{\mu\nu\rho}\partial_\nu\epsilon_{
\rho\alpha\beta} \partial^\alpha A^\beta\epsilon_{\mu\sigma\tau}\partial^\sigma
A^\tau-\frac{1}{2m}\epsilon^{\mu\nu\rho}\partial_\nu
A_\rho\epsilon_{\mu\sigma\tau}\partial^\sigma A^\tau>\ .
\end{equation}
The curved arrow in  Fig.\ref{tree} between $I_0$ and $I^{3th}$  points out that they are also related by the covariant, but not canonical change of variables, $a(A)$  defined by (\ref{amu}).   Due to the higher order derivatives present in the actions none of these
procedures allow to establish the canonical equivalence between the actions
considered and in fact  $I^{3th}$ and  $I^{TM}_2$ are not equivalent 
(hence, the corresponding arrow in the graph above is dotted). 
The model defined by $I^{3th}$ is not unitary as was shown in \cite{DesSJ1999}
where the propagation of  a massive transverse
mode and a spurious massless  ghost was demonstrated. Since the self dual action depends on the longitudinal part of
$a_\mu$ perhaps it is not surprising that the 
change of variables (\ref{amu}), which only define the transverse part of the
field $a$ does not generate a unitary model. 
In  section (\ref{Nonlocal}) we show, using the Hamiltonian analysis, that instead
there exists a non local unitary model which is equivalent to $I^{TM}_2$ and hence to
$I^{TM}$ and $I_{0}$.

The  models related to $I_2$ do not exhaust the  set of master
actions of potential interest. As already
said, all the actions $I_N$  defined by (\ref{IN})
are equivalent to $I_0$. We now check if they  generate other unitary equivalent actions.  Consider then $I_3$ which renaming the fields  takes the form,
\begin{eqnarray}\label{I3}
   I_3 [a,A,B,C]=<-\frac{m}{2}a_\mu a^{\mu}+a_\mu
\epsilon^{\mu\nu\rho}\partial_\nu A_\rho -A_\mu
\epsilon^{\mu\nu\rho}\partial_\nu B_\rho\nonumber\\+B_\mu
\epsilon^{\mu\nu\rho}\partial_\nu C_\rho
-\frac{1}{2} C_\mu
\epsilon^{\mu\nu\rho}\partial_\nu C_\rho>\ \ .
\end{eqnarray}

The action  $I_{FP}^{3th}$ is obtained by unfolding the Maxwell term of $I^{3th}$ with the aid of an auxiliary field, and reads,
\begin{eqnarray}\label{3-FP}
 I^{3th}_{FP}[a,A]&=&<\frac{1}{2}a^\mu a_\nu -
 a^\mu\epsilon_{\mu\alpha\beta}\partial^\alpha A^\beta\\ \nonumber &+&
\frac{1}{2m^2}\epsilon^{\mu\nu\rho}\partial_\nu\epsilon_{
\rho\alpha\beta} \partial^\alpha A^\beta\epsilon_{\mu\sigma\tau}\partial^\sigma\ .
A^\tau>
\end{eqnarray}
  
 The action $I^{3th}_2$ is obtained eliminating $B$ in $I_3^{TM}[A,B,C]$ using the equations of motion. It is given by
\begin{eqnarray}\label{3-2}
 I^{3th}_2[A,C]&=&<-\frac{1}{2m}\epsilon^{\mu\nu\rho}\partial_\nu
A_\rho\epsilon_{\mu\sigma\tau}\partial^\sigma A^\tau\\ \nonumber &+&
\frac{1}{m}\epsilon^{\mu\nu\rho}\partial_\nu C_\rho\epsilon_{\rho\sigma\tau}
\partial^\sigma A^\tau-\frac{1}{2}C_\mu\epsilon^{\mu\nu\rho}\partial_\nu
C_\rho>\ .
\end{eqnarray}
Note that the coupling here is second order. By resolving $C$ in terms of $A$
using the equations of motion in (\ref{3-2})
one obtains $ I^{3th}$, but no fourth order
model may be generated from $I^{3th}_2$ in any evident way. 

Instead, insisting in eliminate $A$ covariantly one recovers $I^{TM}$ without of course
establishing canonical equivalence. The Hamiltonian analysis of $I^{3th}_2$  which we
discuss in the following section shows that it is equivalent to the
ghost propagating $I^{3th}$. Generalizations of $I^{3th}_2$ in the spirit of
(\ref{IN}) and \ref{TMMN}) are straightforward to define and would appear related to $I_N$ with
$N>3$,  but no promising unitary models appear from such analysis.

The action $I^{3th}_2$  may also be obtained using the self-dual change 
of variables (\ref{amu}) in the form $(a\rightarrow a(A),A\rightarrow C)$ in $I_1[a, A]$. Substituting the second field in the same action with a similar change of variables one obtains,
\begin{eqnarray}\label{4-FP}
 I^{4th}_{FP}[a,C]&=&<-\frac{1}{2m}a^\mu a_\nu +
\frac{1}{m}\epsilon_{\mu\sigma\tau}\partial^\sigma a^\tau\epsilon_{\mu\alpha\beta}\partial^\alpha C^\beta\\ \nonumber &+&
\frac{1}{m^2}\epsilon^{\mu\nu\rho}\partial_\nu\epsilon_{
\rho\alpha\beta} \partial^\alpha C^\beta\epsilon_{\mu\sigma\tau}\partial^\sigma
C^\tau>\ .
\end{eqnarray}
By eliminating covariantly $a$ in this action one gets, after some rearrangements,
\begin{equation}\label{4th}
   I^{4th} [C]=\frac{1}{m^{2}}\left(<\frac{1}{2m}\epsilon^{\mu\nu\rho}\partial_\nu
C_\rho\square\epsilon_{\mu\sigma\tau}\partial^\sigma C^\tau-\frac{1}{2} C_\mu
\square\epsilon^{\mu\nu\rho}\partial_\nu C_\rho> \right)\ .
\end{equation}

This fourth order gauge invariant action may also be generated using the self-dual
change of variables $ A(C)$
in the topologically massive action(\ref{TMM}). The action (\ref{4th}) is simply the $TMM$ with an interpolated D'Alembertian operator and propagates ghosts. 
Iterating the process we rise the order of the action by two each time and we
can generate the actions
\begin{equation}\label{2nth}
   I^{2nth} [B_\mu
]=\frac{1}{m^{2(n-1)}}<\frac{1}{2m}\epsilon^{\mu\nu\rho}\partial_\nu
B_\rho\square^{n-1}\epsilon_{\mu\sigma\tau}\partial^\sigma B^\tau-\frac{1}{2}
B_\mu \square^{n-1}\epsilon^{\mu\nu\rho}\partial_\nu B_\rho>\ , 
\end{equation}
with $n=2,3,\ldots$.The TMM may be identified with the case $n=1$

\section{Hamiltonian analysis}
\label{Hamiltonian}
Let us turn to the $(2+1)$ analysis of the dynamics in the so called Hamiltonian
variables \cite{ArnRDM1962,AraCD1980,AraCS1984}. For that consider the  operators
\footnote{We use the metric $\eta=diag\{-1,1,1\}$,
$\epsilon^{012}=-\epsilon_{012}=1$, $\epsilon^{0ij}=\epsilon_{ij}$,
$\epsilon_{ij}\epsilon_{ik}=\delta_{jk}$}
\begin{equation}\label{rho}
   \rho\equiv\sqrt{-\partial_i \partial_i},\ \ \ \rho_i\equiv
\frac{\partial_i}{\rho},\ \ \ \sigma_i\equiv \epsilon_{ij}\rho_j\ \ ,
\end{equation}
which satisfy
\begin{equation}\label{algebra}
\rho_i\rho_i=-1=\sigma_i\sigma_i\ ,\ \epsilon_{ij}\sigma_j=-\rho_i\ ,\
\rho_i\sigma_i=0\ ,\  \square=-\partial_0\partial_0-\rho^2\ \ ,
\end{equation}
and define transverse and longitudinal variables of a vector field $A$ by
\cite{AraCD1980, AraCS1984},
\begin{equation}\label{decomposition}
A_\mu\longrightarrow A_0\ ,\ A_i= -\sigma_iA^T-\rho_iA^L\ ,\ A^T=\sigma_iA_i\ ,\
A^L=\rho_iA_i\ \ .
\end{equation}
The transverse variable $A^T$ is gauge invariant. Define also the auxiliary
gauge invariant variable
\begin{equation}\label{FA}
F_A=\rho A_0+\partial_0A^L\ \ .
\end{equation}
Using these variables the self dual action reduces to
\begin{equation}\label{Selfdualh}
 I_0[a]=\frac{1}{2m}<a_0a_0-a^Ta^T-a^La^L>-<a^TF_a>\ \ ,
\end{equation}
whereas the topologically massive model has the form \cite{DesSJT1982}
\begin{equation}\label{TMMh}
 I^{TM}[A]=\frac{1}{2m}<A^T(\square)A^T+F_AF_A>+<A^TF_A>\ \ .
\end{equation}
Here we see clearly that both  actions propagate only a massive mode and that it
is the coupling of the transverse mode to the gauge invariant variable $F_A$
what generates the mass of the systems. Although $F_A$ includes a time
derivative in its definition it may be verified, by looking at the equations
generated by $A_0$ and $A^L$ separately, that it can be used safely  as a
fundamental variable in this case. This occurs because the variations with respect
to $A_0$ generate a constraint and the variations with respect to $A^L$ generate the
time derivative of this constraint. In some of the models with higher
derivatives discussed below this does not happen. 

For the master action $I_1$, one has
\begin{equation}\label{I1h}
 I_1[a,A]=\frac{1}{2m}<a_0a_0-a^Ta^T-a^La^L>+<A^TF_A-A^TF_a-a^TF_A>\ \ ,
\end{equation}
which reduces directly to (\ref{Selfdualh}) or (\ref{TMMh}) when either one or
the other of the fields is eliminated.
The reduced form of  $I_2[a,A,B]$ and $I^{TM}_2[A,B]$ are  given by
\begin{eqnarray}
 I_2[a,A]&=&\frac{1}{2m}<a_0a_0-a^Ta^T-a^La^L>\nonumber\\
&+&<-A^TF_a-a^TF_A+A^TF_B+B^TF_A-B^TF_B>\ \ ,
\end{eqnarray}
\begin{equation}\label{TMM2h}
 I^{TM}_2[A,B]=\frac{1}{2m}<A^T(\square)A^T+F_AF_A>+<A^TF_B+B^TF_A-B^TF_B>\ \ .
\end{equation}
Eliminating $B^T$ and $F_B$ from  $I_2[a,A,B]$ one recovers the form (\ref{I1h})
of $I_1[a,A]$. Alternatively substituting first the expressions for $a_0$,$a^L$
and $a^T$ one gets the form (\ref{TMM2h}) of $I^{TM}_2[A,B]$. This completes the demonstration of the canonical 
equivalence of the unitary models considered until now.

There is  an alternative approach which can be taken to analyze the dynamical content of 
 $I^{TM}_2[A,B]$ and is to  eliminate $B$ in  (\ref{TMM2h}).
Then, one arrives in one step
to (\ref{TMMh}).  Restricting the functional space adequately in order to have
the inverse operators well defined and  eliminating $A$  in the same action  one
obtains the new equivalent action
\begin{equation}\label{NL1}
 I^{TM}_2[A(B),B]=-\frac{m}{2}<B^TB^T+F_B(\frac{1}{\square})F_B>-<B^TF_B>\ \ ,
\end{equation}
which, in spite of being non local, should propagate a single massive excitation.  This
is enforced by the  equation of motion for $B^T$,
\begin{equation}
 m^2\frac{B^T}{\square}-B^T=0
\end{equation}
which is obtained after eliminating $F_B$. Moreover, including a Chern Simons
term for $A$ in (\ref{I2}) or (\ref{TMM2}) in the form mentioned in  the
previous section simply results in a shift  $m\longrightarrow(m+\beta)$ of the
mass. In the next section we consider this action further . 

For the reduced action of $I^{3th}$ (\ref{3th}) which appeared in the covariant
treatment the canonical analysis gives  \cite{DesSJ1999},
\begin{equation}\label{3thh}
I^{3th}[A]=-\frac{1}{2m}<A^T\square A^T+F_AF_A>-\frac{1}{m}<A^T\square F_A>\ \ ,
\end{equation}
which as mentioned in the previous section propagates a massive particle and a
massless ghost.

The action $I^{3th}_2$ in Hamiltonian variables is given by
\begin{equation}\label{3th2h}
I^{3th}_2[A,C]=\frac{1}{m}<-\frac{1}{2}A^T\square A^T+\frac{1}{2}F_AF_A+A^T\square
C^T+F_AF_C
-mC^T F_C>\ \ ,
\end{equation}
which may be reduced to
\begin{equation}\label{3th2hr}
I^{3th}_2[A,C]=\frac{1}{m}<-\frac{1}{2}A^T\square A^T+A^T\square C^T\ .
-\frac{m}{2}C^T C^T>\ \ ,
\end{equation}
This action is equivalent to (\ref{3thh}) after a field redefinition and hence
is not unitary.
Finally for the ghost propagating $I^{4th}$ we have the reduction
\begin{equation}\label{4thh}
 I^{4th}[B]=\frac{1}{2m^2}<\frac{1}{m}
B^T(\square^2)B^T+F_B(\square) F_B>+<B^T(\square)F_B>\ \ .
\end{equation}

\section{The topologically massive non local model}\label{Nonlocal}

Due to the linearity of the reduction to the Hamiltonian variables, comparison of (\ref{NL1}) and
(\ref{3thh}) allows us to recognize that the non local covariant action
\begin{eqnarray}\label{NL2}
 I^{NL}[B]&=&-\frac{1}{2}<m\epsilon^{\mu\nu\rho}\partial_\nu
B_\rho\epsilon_{\mu\sigma\tau}\frac{\partial^\sigma}{\square}
B^\tau>\nonumber\\&+&\frac{1}{2}<\epsilon^{\mu\nu\rho}\partial_\nu\epsilon_{
\rho\alpha\beta}\frac{\partial^\alpha}{\square}
B^\beta\epsilon_{\mu\sigma\tau}\partial^\sigma B^\tau>
\end{eqnarray}
when expressed in the Hamiltonian variables has also the decomposition
(\ref{NL1}), ({\it i.e} $I^{NL}[B_0,B^T,B^L]=I^{TM}_2[B_0,B^T,B^L]$), and
propagates a single massive excitation. For the second term in (\ref{NL2}) we note  the following identity 
\begin{eqnarray}\label{CSNL} 
I^{CSNL}[B,{\beta}]&\equiv&\frac{\beta}{2}<\epsilon^{\mu\nu\rho}
\partial_\nu\epsilon_{\rho\alpha\beta}\frac{\partial^\alpha}{\square}
B^\beta\epsilon_{\mu\sigma\tau}\partial^\sigma B^\tau>\nonumber\\ 
&=&\frac{\beta}{2}<B_\mu\epsilon^{\mu\nu\rho}\partial_\nu B_\rho>\equiv
-I^{CS}[B,{\beta}]\ ,
\end{eqnarray}
where in the second line we have the pure Chern-Simons action $I^{NL}$ may be deduced covariantly from (\ref{TMM2}) by first noting that the equation of motion (\ref{EMTMM2}) implies
\begin{equation}
 \epsilon^{\mu\nu\rho} \partial_\nu
A_\rho=m\epsilon^{\mu\nu\rho}\frac{\partial_\nu}{\square}\epsilon_{
\rho\alpha\beta} \partial^\alpha B^\beta
\end{equation}
and then substituting this result in (\ref{TMM2}). In analogy with (\ref{amu}) this  can be viewed as a higher derivative change of variables.

Re-writing  the first term in (\ref{NL2}) we could consider the slightly more
general action
\begin{eqnarray}\label{NL3}
I^{NL}_{\beta}[B]&=&\frac{m}{4}<F_{\mu\nu}(B)(\frac{1}{\square})F^{\mu\nu}
(B)>+\frac{\beta}{2}< B^\mu\epsilon_{\mu\sigma\tau}\partial^\sigma B^\tau>
\end{eqnarray}
with $F_{\mu\nu}(B)=\partial_\mu B_\nu-\partial_\nu B_\mu$. This is written in
the Hamiltonian variables as,
\begin{equation}\label{betah}
 I^{NL}_{\beta}[B^T,F_B]=-\frac{m}{2}<B^TB^T+F_B(\frac{1}{\square}\ ,
)F_B>-\beta<B^TF_B>\ \ .
\end{equation}
and propagates  a single particle with  modified mass $m/{\beta}$.

To grasp this result in a covariant way we note that  (\ref{NL3}) is gauge
invariant  and may be rewritten in the form
\begin{equation}
I^{NL}_{\beta}[B]=<-\frac{m}{2} B^\mu
B_\mu+\frac{\beta}{2}B^\mu\epsilon_{\mu\sigma\tau}\partial^\sigma B^\tau >
-\frac{m}{2}< \partial_\mu B^\mu\frac{\partial_\nu}{\square}B^\nu > \ \ .
\end{equation}
This is the self-dual model of mass $m/\beta$ plus a non-local term which
vanishes in the gauge $\partial_\mu B^\mu=0$.

\section{The Fierz-Pauli non local model}\label{NonlocalFP}
The natural generalization to the PM of the ideas worked in the previous sections
is to apply the self dual change of variables (\ref{amu}) in (\ref{FP}). This leads to the third order action,
\begin{eqnarray}\label{FPSD}
I^{FPSD}[B]=\frac{1}{4}<F_{\mu\nu}(B)F^{\mu\nu}(B)>-\frac{1}{4m^2}<F_{\mu\nu}
(B){\square}F^{\mu\nu}(B)>
\end{eqnarray}
which propagates a massless ghost together with the two massive modes.
The remaining option is to explore the non-local alternatives. The structure observed so far suggests the following model,
\begin{eqnarray}\label{FPNL}
I^{FPNL}[B]=-\frac{1}{4}<F_{\mu\nu}(B)F^{\mu\nu}(B)>+\frac{m^2}{4}<F_{\mu\nu}
(B)(\frac{1}{\square})F^{\mu\nu}(B)>
\end{eqnarray}
which has been discussed in connexion with Stueckelberg formalism \cite{RueHR2004}. 
The non-local Maxwell term may be viewed as a Fierz-Pauli mass term
plus a non-local gauge fixing term. The Hamiltonian
analysis  establishes that this action may be written in the form
\begin{eqnarray}\label{FPNLh}
I^{FPNL}[B]=\frac{1}{2}<B^T(\square-m^2)B^T>+\frac{1}{2}<F_BF_B-m^2F_B(\frac{1}{
\square})F_B>\ ,
\end{eqnarray}
which suggests the propagation of two massive modes with the same mass. To compare, the Hamiltonian decomposition of the Proca-Fierz-Pauli action after eliminating $A_0$ takes the form
\begin{eqnarray}\label{FPh}
I^{FP}[B]=\frac{1}{2}<A^T(\square-m^2)A^T>+\frac{m^2}{2}<A^L(\frac{\square-m^2}{
\rho^2+m^2})A^L>\ .
\end{eqnarray}

The  non local Maxwell term may be induced  by a $BF$ coupling with a massless 
auxiliary field in a way resembling the mechanism in 2D for the Schwinger model.
In that case \cite{LowJS1971} the non-local Maxwell term of (\ref{FPNL}) which
appears after integrating out  the fermions, is expressed in terms of a local
scalar field which propagates a massive excitation and the original field which
does not propagate. Here  we  introduce a second vector field with a $BF$
coupling to obtain the same behavior. In terms of  the gauge fixed massless
Maxwell action 
\begin{eqnarray}
I^M(C,\kappa)=<-\frac{1}{4}F_{\mu\nu}(C)(F^{\mu\nu}(C)+\frac{\kappa}{2}
\partial_\mu C^\mu\partial_\nu C^\nu>
\end{eqnarray}
which in 3-D propagates a single degree of freedom, the model that we  consider reads
\begin{equation}
I_{CB}=I^M(C,\kappa)+<\frac{1}{2}C_\mu\epsilon^{\mu\nu\rho}\partial_\nu
B_\rho>+I^M(B,\tilde{\kappa})\ .
\end{equation}
This model propagates two massive modes of opposite helicity as can be shown by means of the Hamiltonian analysis or doing the decoupling change of variables
\begin{equation}
B_ \mu = \frac{1}{\sqrt{2}} [A_\mu ^1 + A_\mu ^2 ]\ \ , \ \ C_ \mu = \frac{1}{\sqrt{2}} [A_\mu ^1 - A_\mu ^2 ]\ .
\end{equation}
For the  analysis in the functional integral introduce the operators associated to the gauge fixed Maxwell action 
\begin{eqnarray}
D_{\mu\nu}^{M}(\kappa)=\square\eta_{\mu\nu}-(\kappa+1)\partial_\mu\partial_\nu\
\ \ ,\ \ \
G_{\nu\rho}^{M}(\kappa)=\frac{\eta_{\nu\rho}}{\square}-\frac{(\kappa+1)}{\kappa}
\frac{\partial\nu\partial_\rho}{\square^2}
\end{eqnarray}
 which satisfy
\begin{equation}
 D_\mu^{M\nu}(\kappa)G_\nu^{{M}(\kappa)\rho}=\delta_\mu^{\ \rho}\ \ ,\ \
G^{M\nu}_{\ \rho}(\kappa)\epsilon^{\rho\alpha\beta}\partial_\alpha
C_\beta=\epsilon^{\nu\alpha\beta}\frac{\partial_\alpha}{\square} C_\beta\ .
\end{equation}
Then we have  the identity,
\begin{eqnarray}\label{2MNL}
&\ & \int {\cal D}C{\cal D}B
e^{-[I^M(C,\kappa)+<\frac{1}{2}C_\mu\epsilon^{\mu\nu\rho}\partial_\nu
B_\rho>+I^M(B,\tilde{\kappa})]}\nonumber\\
&=&\int {\cal D}C'{\cal D}B
e^{-[I^M(C',\kappa)+<\frac{m^2}{4}<F_{\mu\nu}(B)(\frac{1}{\square})F^{\mu\nu}
(B)>+I^M(B,\tilde{\kappa})]}\ ,
\end{eqnarray}
where
$C'_\mu=C_\mu-mG_\mu^{M\nu}(\kappa)\epsilon_{\nu\alpha\beta}\partial^\alpha
B^\beta$. 
One of the vector fields is the physical field which couples to the sources and acquires its mass through the appearance of the non-local Maxwell term. The second  field is an auxiliary field which does not couple to the sources and contributes a constant factor to the generating functional which in turn is absorbed in the normalization constant.

Alternatively, the Fierz Pauli non local model emerges from the Stueckelberg model in any dimension 
\begin{equation}\label{IS}
 I^{S}(A,\Phi)=-\frac{1}{4}<F_{\mu\nu}(A)F^{\mu\nu}(A)>-\frac{m^2}{2}<(A_\mu+\frac{1}{m}\partial_\mu\Phi)(A^\mu+\frac{1}{m}\partial^\mu\Phi)>
\end{equation}
by decoupling $\Phi$ in the functional integral in the usual way. Again a factor corresponding to a uncoupled massless scalar field  remains in the generating functional and is absorbed in the normalization constant.

\section{Discussion}

In this paper we presented a unified analysis of massive vector models in $3D$ in terms of a set of master actions related by duality transformations. Using the Hamiltonian $(2+1)$ canonical variables the
equivalence  between the various models was checked explicitly. 
This allowed us to identify a new unitary non local vector model with a single
massive excitation. The  non-local action emerges as an improvement, suggested by the Hamiltonian analysis of a third order non unitary model which appears naturally in the covariant
reduction of one of the first order actions generated by duality. Non-local gauge invariant actions in various dimensions have also been discussed in the literature in relation with the Stueckelberg formalism (see \cite{RueHR2004} for a recent review). It is then of interest that in 3D complete canonical equivalence with a local model may be demonstrated using the Hamiltonian variables. 

The non local terms discussed also may be extracted by decoupling the fields in the path
integral formalism as was illustrated with the non-local Fierz Pauli model and are related in this way to the Stueckelberg field \cite{RueHR2004}. It remains to be explored if interactions could be incorporated consistently to these models. It is not ruled out that the techniques used in this work when applied to an action with higher $N$ in (\ref{IN}) or (\ref{TMMN}) may generate also non local unitary models.

Although the non local formulation may perhaps be only of limited value in an
operatorial quantum mechanical description it should be useful in the path
integral framework in connection with the dual representation of low dimensional
systems. In particular the system defined by (\ref{NL3})  may be also of
interest in relation with  bosonization of fermion fields in 3D
\cite{MarE1991,MarES1992,KovAK1993,BurLQ1994,FraES1994}. In 3D the bosonized
fermion current is identified with a Chern Simons topological current
\cite{MarES1992}.
\begin{equation}
 \bar\psi\gamma^\mu\psi\sim\epsilon^{\mu\nu\rho}\partial_\mu A_\rho\ \ .
\end{equation}
Both in the operatorial \cite{MarE1991} and in the functional approaches
\cite{BurLQ1994,FraES1994} the effective bosonized action is shown in general to
be  non local  and it could be interesting to determine if there is some
relation with the system at hand.

Some of the ideas presented here may be extended to the self-dual, second order \cite{AraCK1986},
topologically massive \cite{DesSJT1982} and fourth order \cite{BerEHT2009,DalDM2009} three
dimensional massive gravity models which share a part of the structure of the vector models  and also display  duality relations.

\section{Acknowledgments}

This work was supported by DID-USB GID-30. JS thanks the members of  Centro de
F\'{\i}sica Fundamental at ULA for hospitality.

\appendix
\numberwithin{equation}{section}
\section{Chern Simons self duality}

We can get more insight about where the non-local terms come from by considering the dual to
the pure Chern Simons action. As mentioned in the introduction this is obtained from the gauge fixed
Chern Simon action of the gauge field $A$ by means of $BF$ coupling with a new
vector field $B$ \cite{SteJ1997},
\begin{eqnarray}
 I_{gf}^{CSD}[A,B]&=&<-\frac{\beta}{2}A_\mu\epsilon^{\mu\nu\rho}\partial_\nu A_\rho+\frac{
\kappa}{2}\partial_\mu A^\mu\partial_\nu A^\nu\nonumber\\&-&
A_\mu\epsilon^{\mu\nu\rho}\partial_\nu B_\rho
+\frac{\tilde\kappa}{2}\partial_\mu B^\mu\partial_\nu B^\nu>
\end{eqnarray}
where the subscript ${gf}$ is a remainder that a gauge fixed term has been
include for each field. 
Introduce now,
 \begin{equation}
 D^{\mu\nu}=\beta\epsilon^{\mu\nu\rho}
\partial_\rho-\kappa\partial^\mu\partial^\nu,\ \ \
G_{\nu\rho}=\frac{1}{\beta}\epsilon_{\nu\rho\tau}\frac{\partial^\tau}{\square}
-\frac{1}{\kappa}\frac{\partial_\nu\partial_\rho}{\square^2}\ \ ,
\end{equation}
which satisfy in symbolic Heaviside notation
\begin{equation}
D^{\mu\nu}G_{\nu\rho}=\delta^\mu_\nu\ \ ,\ \
G_{\nu\rho}\epsilon^{\rho\tau\sigma}\partial_\tau
B_\sigma=\frac{1}{\beta}\epsilon_{\nu\rho\lambda}\frac{\partial^\lambda}{\square
}\epsilon^{\rho\tau\sigma}\partial_\tau B_\sigma
\end{equation}
and consider the vacuum amplitude
\begin{equation}
 Z^{CSD}[0]=\int {\cal D}A{\cal D}B e^{-I_{gf}^{CSD}[A,B]}\ .
\end{equation}
Then shifting $A_\mu\longrightarrow
A_\mu+G_{\mu\nu}\epsilon^{\nu\tau\sigma}\partial_\tau B_\sigma$ in the
generating functional after completing the square in the action one gets
\begin{equation}
 Z^{CSD}[0]=\int {\cal D}A e^{-I_{gf}^{CS}[A,\beta]}\int {\cal D}B
e^{-I_{gf}^{CSNL}[B,\tilde{\beta}]}
\end{equation}
where  $\tilde{\beta}=-\frac{1}{\beta}$. This shows that  the action
$I^{CSNL}=-I^{CS}$ in equation (\ref{CSNL}) is the dual to the pure Chern Simons
model. The appearance of the apparently non-local term in $I_{gf}^{CSNL}$ may be
traced to the term  
$-\frac{1}{\beta}\epsilon_{\nu\rho\tau}\frac{\partial^\tau}{\square}$ in the
covariant propagator.

\end{document}